\documentclass{article}

\usepackage[english]{babel}

\usepackage[letterpaper,top=2cm,bottom=2cm,left=3cm,right=3cm,marginparwidth=1.75cm]{geometry}

\usepackage{amsmath}
\usepackage{graphicx}
\usepackage[colorlinks=true, allcolors=blue]{hyperref}
\usepackage{tabularx} 
\usepackage{booktabs}  
\usepackage{authblk}

\begin{document}
\title{A Unified Critical Scaling Theory for Macroscopic Lightning and Quantum Avalanches: From Three-Dimensional Directed Percolation to Testable Predictions}
\author{Zhe Li}
\affil{China Electric Power Research Institute, Beijing, China; lizhe5@epri.sgcc.com.cn}

\maketitle

\begin{abstract}
Lightning, the most colossal discharge in nature, and flux avalanches in quantum superconductors, which are phenomena separated by twenty orders of magnitude in scale, display striking fractal similarity. We demonstrate that this is no mere analogy but reveals a deep physical unity. By mapping both onto a three-dimensional reaction-diffusion-advection equation from non-equilibrium statistical physics, we show they belong to the same critical universality class: three-dimensional Directed Percolation (3D-DP). This yields a unified set of universal critical exponents (e.g., avalanche size distribution exponent $\tau \approx 1.41$, fractal dimension $D_f \approx 2.5$) for both systems. Furthermore, by incorporating the anisotropy and turbulence coupling intrinsic to real thunderstorm environments, we predict novel effects such as anisotropic fractality of lightning channels and the systematic shift of critical exponents by turbulence. The core theoretical breakthrough lies in proposing a geometric correspondence of quantum phase information: through a rigorous mapping, the microscopic quantum phase coherence of superconductors is translated into the curvature and torsion distributions of macroscopic lightning channels, revealing a quantum statistical fingerprint emergent in classical geometry. This framework not only provides a unified paradigm for understanding dissipative structures across scales but also, via seven testable predictions, opens avenues for simulating natural lightning with laboratory quantum systems and developing novel physical early-warning methods.
\end{abstract}

\section{Introduction}

\begin{figure}
    \centering
    \includegraphics[width=0.75\linewidth]{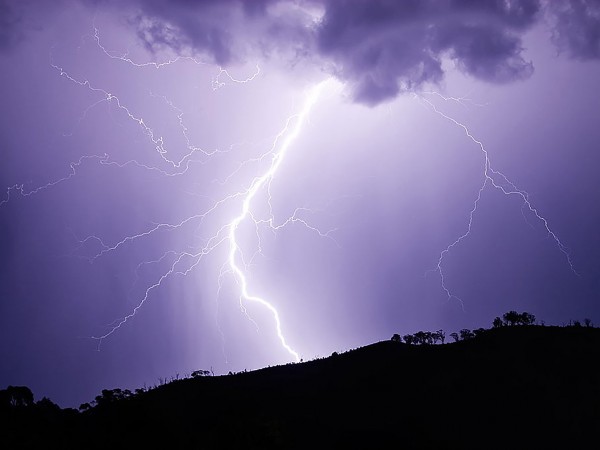}
    \caption{Cloud-to-ground lightning}
    \label{fig:1}
\end{figure}

\begin{figure}
    \centering
    \includegraphics[width=0.75\linewidth]{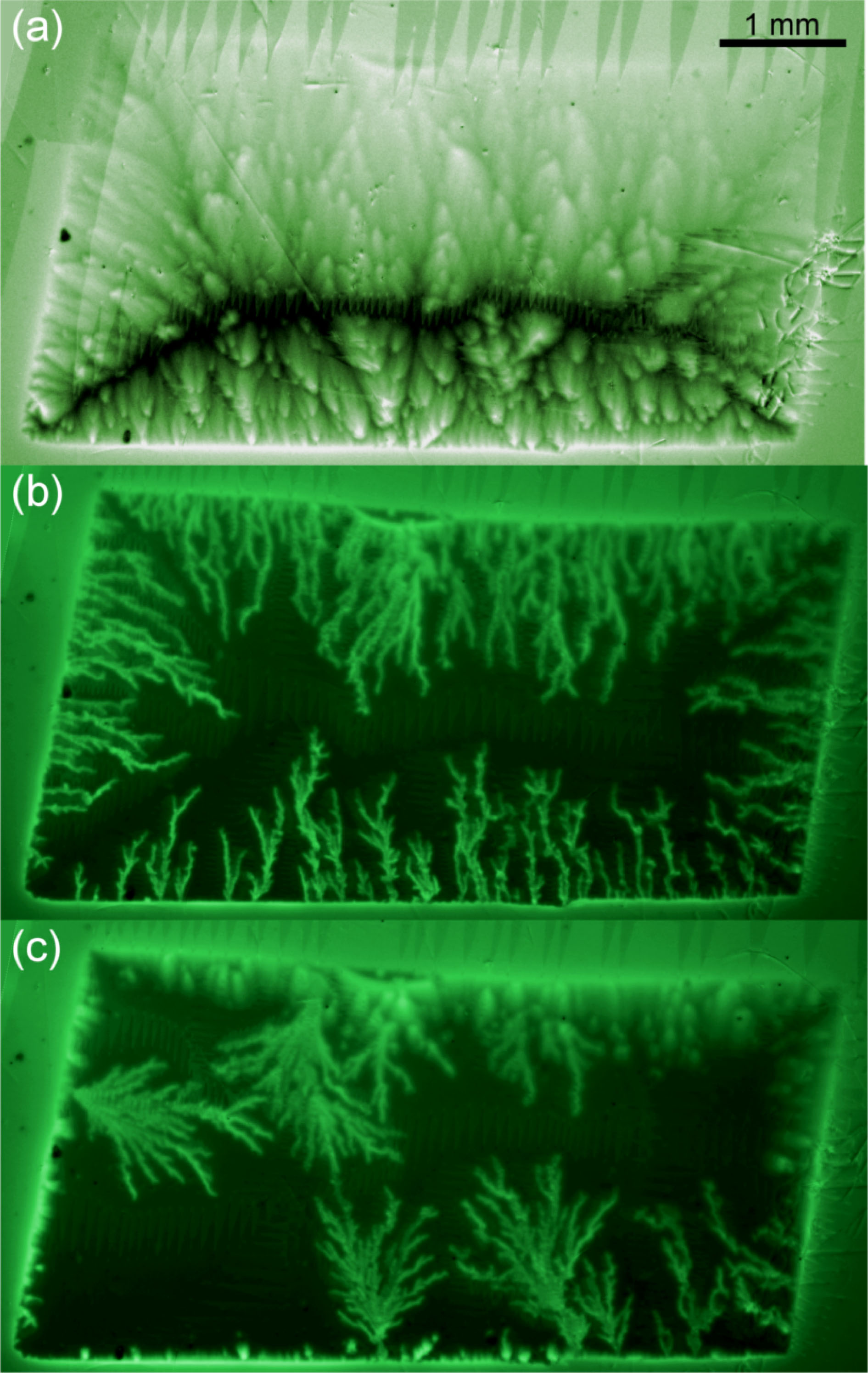}
    \caption{Magneto-optical(MO) images taken after a ZFC procedure, applying \textit{\textit{H}}=46Oe:(a) full penetration of the ﬂux at 12K, where the granular character of the ﬁlm can be clearly observed.(b) Flux jumps imprinted in V\textsubscript{\textsubscript{3}}Si thin ﬁlm at 2.47K and (c) at 7.0K \cite{Pinheiro2019}.}
    \label{fig:2}
\end{figure}

Lightning fig:\ref{fig:1} and superconducting flux avalanches fig:\ref{fig:2}, residing in the distinct fields of atmospheric electricity and condensed matter physics, respectively, share complex dendritic patterns. Is this morphological similarity across macro- and mesoscopic scales coincidental, or does it reveal a deeper physical unity in non-equilibrium dissipative systems? Traditionally studied in isolation: the fractal dimension of lightning channels has been extensively measured and linked to turbulent activity, while superconducting avalanches have been confirmed as a canonical example of the Directed Percolation (DP) universality class \cite{Field1995,bassler1998simple}. A fundamental question arises: Can a unified mathematical framework be established, describing both phenomena with the same set of universal laws \cite{Tinkham1996,Rakov2003}?

Recent breakthroughs in related fields have set the stage for a deeper unification. In condensed matter physics, experiments on \textit{interdependent superconducting networks} have revealed cascading phase transitions mediated by spontaneous critical processes, providing a physical paradigm for avalanche dynamics in complex coupled systems \cite{Bonamassa2023Interdependent}. Concurrently, the observation of \textit{bath-induced quantum avalanches} in many-body localized systems underscores the unique propagation mechanisms of avalanches in the quantum realm \cite{Leonard2023Signatures}. Recent theoretical work further elucidates the conditions for thermal avalanche propagation in disordered quantum spin systems \cite{Szoldra2024CatchingAvalanches}. Furthermore, advances in quantum simulation have enabled the characterization of non-equilibrium phase transitions, such as the quantum extension of the classical contact process (directed percolation universality class), on programmable quantum processors \cite{Chertkov2023Characterizing}. These advances, from the perspectives of \textit{complex networks}, \textit{quantum many-body systems}, and \textit{quantum simulation}, have significantly deepened our understanding of avalanches and critical phenomena at disparate scales.

Critically, new observational insights bridge these concepts to the natural world. The latest research finds a strong correlation between the morphological complexity of lightning channels and in-cloud turbulent kinetic energy dissipation rate, hinting at the crucial modulation of discharge dynamics by environmental spatiotemporal noise \cite{Li2024}. This work aims to synthesize these threads. We demonstrate that under a coarse-grained description, lightning and superconducting avalanches obey the same effective field theory, both belonging to the three-dimensional Directed Percolation (3D-DP) universality class. Beyond this scaling unification, we advance a revolutionary viewpoint: the microscopic quantum phase information of superconductors can statistically "emerge" in the geometric structure of macroscopic lightning. Here, we establish this unified framework and present a set of quantitative, testable predictions that bridge these scales.This lays the groundwork for the emerging interdisciplinary field of "quantum atmospheric electricity."

\section{A Three-Dimensional Unified Theoretical Framework}

\subsection{Mapping of Physical Quantities}

At a coarse-grained level focusing on charge carrier density-electric field coupling, we establish the following core mapping tab:\ref{tab:1}.
The lattice model description of superconducting vortex avalanches provides a theoretical foundation for this mapping \cite{bassler1998simple}.
\begin{table}[ht]
\centering
\small
\setlength{\tabcolsep}{3pt}
\begin{tabular}{p{2.7cm}p{2.7cm}p{2.7cm}p{2.7cm}}
\toprule
\textbf{Physical System/Quantity} & \textbf{Superconducting Avalanche (3D)} & \textbf{Atmospheric Discharge (3D)} & \textbf{Mapping Relation} \\
\midrule
Order Parameter & Magnetic Flux Density $B(\mathbf{r},t)$ & Electron Density $n(\mathbf{r},t)$ & 3D Scalar Field \\
Current Density & Supercurrent Density $\mathbf{J}_s$ & Electron Current Density $\mathbf{J}_e$ & 3D Vector Field ($\nabla \cdot \mathbf{J} \approx 0$) \\
Topological Defect & Flux Vortex Line & Lightning Channel (Streamer) & 3D Curve \\
Topological Evolution & Vortex Branching/Merging & Streamer Branching/Coalescence & 3D Topological Process \\
Spatial Disorder & Pinning Center Distribution & Aerosol/Electronegative Regions & System Inhomogeneity \\
\bottomrule
\end{tabular}
\caption{Mapping of Physical Quantities between Superconducting Avalanches and Atmospheric Discharge}
\label{tab:1}
\end{table}

\subsection{Three-Dimensional Evolution Equation}

The dynamics of the above systems can be described by an equation for a generalized order parameter  \(u(\mathbf{r}, t)\) (proportional to charge carrier density):

\begin{equation}
\begin{split}
\frac{\partial u}{\partial t} &= \nabla_{\perp} \cdot \left( D_{\perp}(u) \nabla_{\perp} u \right) 
                               + \frac{\partial}{\partial z} \left( D_{\parallel}(u) \frac{\partial u}{\partial z} \right) \\
                              &\quad - \nabla \cdot (u \mathbf{v}) + f(u, \nabla u) - \eta(\mathbf{r}, t)
\end{split}
\end{equation}
where  \(D_{\perp}\) and  \(D_{\parallel}\) are anisotropic diffusion coefficients,  v is a convection velocity (e.g., updraft), \(f(u) = r u - g u^{2} + \cdots\) is a source term expanded near the critical point, and \(\eta\) is a multiplicative noise satisfying \(\langle \eta(\mathbf{r}, t) \, \eta(\mathbf{r}', t') \rangle \propto u(\mathbf{r}, t) \; \delta(\mathbf{r} - \mathbf{r}') \, \delta(t - t')
    \label{eq:noise_correlation}\)—a key feature of the DP universality class.Varying the action yields the nonlinear Schrödinger equation (Gross-Pitaevskii equation):

\section{Critical Behavior: The 3D Directed Percolation (3D-DP) Universality Class}

Equation (1) is the standard form for systems with an absorbing-state phase transition, belonging to the \textbf{three-dimensional Directed Percolation (3D-DP)} universality class \cite{Hinrichsen2000,Munoz1998}. Via renormalization group theory and series expansion, we obtain its critical exponents \cite{lubeck2005scaling,deoliveira2011quasi}:
\begin{table}[h]
\centering
\small 
\setlength{\tabcolsep}{3pt} 
\begin{tabular}{cccp{6cm}} 
\toprule
Exponent & Symbol & 3D Value & Physical Meaning\\
\midrule
Order Parameter & $\beta$ & $\approx 0.81$ & Steady-state active density: $\rho \sim (p - p_c)^{\beta}$\\

Spatial Correlation & $\nu_{\perp}$ & $\approx 0.58$ & Spatial correlation length: $\xi_{\perp} \sim |p - p_c|^{-\nu_{\perp}}$\\

Temporal Correlation & $\nu_{\parallel}$ & $\approx 1.11$ & Temporal correlation length: $\xi_{\parallel} \sim |p - p_c|^{-\nu_{\parallel}}$\\

Dynamic Exponent & $z = \nu_{\parallel} / \nu_{\perp}$ & $\approx 1.9$ & Space-time scaling: $t \sim x^{z}$\\

Avalanche Size Dist. & $\tau$ & $\approx 1.41$ & Avalanche size distribution: $P(s) \sim s^{-\tau}$ (Core Prediction 1)\\

Fractal Dimension & $D_f$ & $\approx 2.5$ & Geometric fill of a single avalanche cluster in 3D space (Core Prediction 2)\\
\bottomrule
\end{tabular}
\caption{Critical Exponents}
\label{tab:2}
\end{table}
Core Conclusion: Regardless of microscopic mechanisms, the avalanche statistics, spatiotemporal correlations, and fractal structure of macroscopic lightning and superconducting avalanches near the critical point are governed by this identical set of universal exponents \cite{Hinrichsen2000,Takeuchi2007}.

\section{Geometric Correspondence of Quantum Phase Information}

A profound question arises: Does the quantum phase coherence of superconductors have a counterpart in macroscopic lightning? We propose that the three-dimensional geometric structure (curvature $\kappa$, torsion $\tau$) of a lightning channel encodes equivalent "phase" information.

\subsection{Geometric Mapping}

The phase gradient $\nabla\theta$ of the superconducting order parameter defines the superfluid velocity direction. We construct a geometric mapping $\Phi: \theta \mapsto (\kappa, \tau)$ that translates the information of the quantum phase field $\theta(\mathbf{r})$ around a superconducting vortex line into the classical geometric quantities---curvature $\kappa$ and torsion $\tau$---of the corresponding space curve. This mapping is conceptually rooted in the modern theory of geometric phases and quantum geometry \cite{berry1984quantal,xiao2010berry}, which establishes a profound connection between the topology of quantum states and their emergent geometric properties. (See Supplementary Material S1 for derivation):
\begin{align*}
\kappa_c(s) = \mathcal{F}[\theta] &= \lim_{R \to 0} \frac{1}{|\nabla_\perp\theta|} |\nabla_\perp^2\theta|, \\
\tau_c(s) = \mathcal{G}[\theta] &= \frac{\hbar}{2m} \frac{\oint_{C(R)} (\nabla\theta \times \nabla^2\theta) \cdot d\mathbf{A}}{\oint_{C(R)} |\nabla_\perp\theta|^2 dl}
\end{align*}
where $s$ is the arc length. This mapping correlates topological defects of the phase vortices with branching points of the channel. Recent studies on geometric induction in quantum vortices  \cite{Jiang2022} provide tangential theoretical support for the physical plausibility of this mapping.

\subsection{Scale Invariance and Universal Statistics}

At the 3D-DP critical point, this mapping exhibits scale invariance. The scaling exponent for the geometric quantities is uniquely determined by the universal critical exponents: $\alpha = 1/(\nu_{\perp z}) \approx 0.91$. Consequently, the curvature and torsion of lightning channels and superconducting vortex lines should obey the same joint statistical distribution:
\[
P(\kappa, \tau; L) = L^{2\alpha} f\left( \frac{\kappa}{L^\alpha}, \frac{\tau}{L^\alpha} \right)
\]
where $f$ is a universal scaling function.

\subsection{Physical Interpretation: An Emergent Classical Fingerprint}
This correspondence does not imply that macroscopic lightning exhibits quantum coherence. Instead, it reveals a profound connection: the microscopic quantum phase information of a superconductor leaves a statistical imprint on the geometry of a macroscopic lightning channel, much like a fingerprint pressed into clay. In this sense, the lightning channel acts as a classical recorder, encoding the statistical signatures of quantum critical fluctuations into its enduring geometric form. The topological constraints and correlation patterns inherent in the quantum phase field, by virtue of the shared critical dynamics (the 3D-DP universality class), are ``frozen in'' and statistically manifested as the winding and branching of the classical discharge path. Thus, the meandering of a lightning channel can be viewed as the emergent geometric fingerprint of an underlying quantum statistical order, now observable in the classical world.

\section{Testable New Predictions}

Based on the unified framework, we propose seven quantitative predictions testable in three dimensions:
\begin{enumerate}
    \item \textbf{3D Critical Exponent Consistency:} The avalanche size distribution exponent $\tau$ extracted from 3D radiation source data (e.g., LMA) of natural lightning should match that of flux avalanches in 3D superconductors, both being $\tau \approx 1.41 \pm 0.1$  \cite{Field1995,Altshuler2004}.

    \item \textbf{Anisotropic Fractal Dimensions:} The vertical ($D_f^{\parallel}$) and horizontal projection ($D_f^{\perp}$) fractal dimensions of lightning channels satisfy: $D_f^{\parallel}/D_f^{\perp} \approx 1.9$.

    \item \textbf{Turbulence-Modulated Critical Behavior:} The avalanche exponent for lightning will exhibit a systematic shift in storms with different turbulent kinetic energy dissipation rates (EDR, $\varepsilon$): $\tau(\varepsilon) = \tau_0 + \alpha\varepsilon^\beta$, where $\tau_0$ is the pure DP value. Recent observations provide qualitative support  \cite{Li2024}.

    \item \textbf{Finite-Size Scaling:} The size distribution of lightning radiation source clusters follows $P(s) \sim s^{-\tau} \exp(-s/s_0)$, where $s_0 \propto H^{D_f}$ and $H$ is the charge layer thickness.

    \item \textbf{Power-Law Curvature-Torsion Correlation:} The spatial correlation between local curvature $\kappa$ and torsion $\tau$ along a channel satisfies $\langle \kappa(s)\tau(s+r) \rangle \sim r^{\gamma_{\kappa\tau}}$, with $\gamma_{\kappa\tau} = 2/(\nu_{\perp z}) - 3 \approx -1.18$.

    \item \textbf{Singularity at Branching Points:} The local curvature distribution near channel branching points follows $P(\kappa |_{\text{branch}}) \sim \kappa^{-\mu_b}$, with exponent $\mu_b$ larger than the bulk value.

    \item \textbf{Critical Condition for ``Quantum Lightning Rods'':} Using a Rydberg atom ensemble as a tunable discharge seed, the breakdown threshold field $E_c$ and ensemble size $R$ satisfy $E_c(R) = E_c^{\infty} + (E_0 - E_c^{\infty}) \exp(-R/\xi)$, where the correlation length $\xi$ obeys DP scaling.
\end{enumerate}

\section{Dialogue with Existing Observations}

Our theory aligns deeply with recent precise observations and provides a unified explanation.
\begin{enumerate}
\item \textbf{Benchmark for Fractal Dimension:} 
Numerous measurements report the 2D projection fractal dimension $D_{2D}$ of negative cloud-to-ground lightning to be $\sim 1.2\text{--}1.3$ \cite{Amarasinghe2015,antrias2024fractal,Matsui2002}, consistent in order of magnitude with the 2D projection ($D_{2D} = D_{f} - 1 \approx 1.5$) derived from our 3D theoretical value $D_{f} \approx 2.5$. It is crucial to emphasize that this theoretical value of $D_{f} \approx 2.5$ is derived for an isotropic, ideal three-dimensional system at the critical point. Deviations likely stem from anisotropy and finite-size effects, which are inherently described by our framework. In real atmospheric discharges, strong inherent anisotropies---such as the directional bias imposed by the vertical gravitational field and the large-scale electric field, as well as the finite thickness of charge layers---can significantly modify the effective fractal dimension measured in observations. Therefore, the comparison between the isotropic 3D-DP theoretical prediction and actual measurements must account for these anisotropic corrections, which our framework explicitly incorporates through the anisotropic scaling relations (e.g., Prediction 2).
\item \textbf{Direct Evidence for Turbulence Modulation:} Li et al. (2024)  \cite{Li2024} quantitatively demonstrated a significant spatial correlation between lightning channel complexity and EDR, providing direct observational support for our Prediction 3, indicating that turbulence acts as a strong noise source modulating the critical dynamics.

\item \textbf{Spatial Leader and Avalanche Cluster Imagery:} The observed formation and connection of "space leader" clusters ahead of a negative leader's head in high-speed photography  \cite{Qi2017} aligns remarkably with the dynamical image of "active sites" randomly "infecting" neighbors in the DP model. This provides the micro-physical basis for viewing lightning channel development as a critical avalanche cluster.
\end{enumerate}

\section{Discussion and Outlook}

\subsection*{Significance of the Unified Picture}

This work places a macroscopic natural phenomenon and a mesoscopic quantum process under the same critical scaling theory for the first time, revealing a profound unity in laws across scales in non-equilibrium systems. It moves beyond phenomenological analogy to establish a quantitative, testable mathematical framework.

\subsection*{Interdisciplinary Extension: Implications for Neural Systems}

It is noteworthy that the Directed Percolation (DP) universality class underlying our unified framework is also widely used to describe the propagation dynamics of neuronal avalanches in the cerebral cortex  \cite{Beggs2003,Friedman2012}. This suggests that macroscopic lightning, mesoscopic quantum avalanches, and microscopic neural population activity may share the same fundamental physics of non-equilibrium critical propagation. A fascinating prospect is whether the geometric scaling laws predicted by our theory (e.g., power-law distributions of curvature/torsion) could also be detected in the spatiotemporal patterns of neuronal avalanches. This opens new possibilities for testing critical geometric theories in the living brain  \cite{Ponce2025}.

\subsection*{Interdisciplinary Extension: The Macroclimatic Driving Force of Lightning and Its Correlation with Mesoscopic Critical Dynamics}

The critical avalanche dynamics of lightning channel formation revealed in this work provides a microphysical basis for understanding the statistical regularities of lightning activity at larger scales. It is noteworthy that Romps et al., based on climate models proposed that the lightning flash rate ($F$) is proportional to the product of Convective Available Potential Energy (CAPE) and precipitation rate ($P$) ($F \propto \text{CAPE} \times P$)  \cite{Romps2014}. Within the framework presented here, $\text{CAPE} \times P$ can be interpreted as a macro-environmental proxy variable that modulates the 'infection probability' $\rho$ in the three-dimensional directed percolation (3D-DP) process. CAPE provides the potential energy driving the discharge, while $P$ reflects the abundance of electrification substances; together, they determine the propensity of the system to deviate from the absorbing state and approach the critical point. This qualitative correspondence suggests that the projected increase in lightning frequency under global warming may fundamentally arise from climatic forcings that drive the atmospheric electrical system more frequently into or near its critical regime. A promising future direction is to combine high-resolution climate models with the geometric scaling laws developed herein, to predict how not only the number of lightning flashes but also the morphology and energy statistics of individual flashes will systematically evolve under a warming climate.

\subsection*{Dialogue with Contemporary Advances}
The three-dimensional DP unified framework established here resonates deeply with several recent advances highlighted in \textit{Nature Physics}. First, the universal avalanche statistics predicted for both lightning channels and flux jumps provide an extreme-scale counterpart---spanning natural and quantum material systems---to the study of \textbf{cascading failures in interdependent networks}\cite{Bonamassa2023Interdependent}. Our framework thus provides a contrasting physical realization, demonstrating how the same universal class (3D-DP) governs cascades from engineered quantum networks to atmospheric giants. Second, our proposed concept of a \textit{geometric correspondence of quantum phase information} suggests a novel avenue for characterizing the recently observed \textbf{quantum avalanche propagation}\cite{Leonard2023Signatures}: the spatiotemporal path of an avalanche may encode geometric fingerprints of its microscopic quantum origins. Our mapping $\Phi$ provides a concrete theoretical tool to decode such potential geometric fingerprints from avalanche data. Finally, the specific theoretical predictions of this work, particularly the precise critical exponents and finite-size scaling relations, offer well-defined and testable targets for future \textbf{verification of non-equilibrium DP transitions on quantum simulators}, such as Rydberg atom arrays or superconducting quantum processors\cite{Chertkov2023Characterizing}. This bridges our theoretical predictions with the most advanced experimental platforms, enabling a direct quantum simulation of the atmospheric-scale critical phenomena we describe. Connecting laboratory-controllable quantum critical systems with the uncontrollable yet observable giant system of nature (lightning) through a common mathematical language is the core vision of the emerging interdisciplinary field of ``quantum atmospheric electricity.''

\subsection*{Future Directions}

\begin{enumerate}
\item \textbf{Experimental Verification:}  Systematically test the predictions through collaborations between lightning observation and superconducting experiment teams.

\item \textbf{Theoretical Extension:}  Apply the framework to other discharge phenomena (e.g., electrostatic discharge, dielectric barrier discharge).

\item \textbf{Applied Exploration:} Develop novel lightning early-warning algorithms based on critical exponents; protect superconducting devices from flux avalanches by drawing inspiration from lightning mechanisms.
\end{enumerate}

\section{Summary}

\begin{enumerate}
\item \textbf{Theoretical Derivation:}  The evolution equation is derived from the time-dependent Ginzburg-Landau equation and streamer discharge models via adiabatic elimination of fast variables. Critical exponents are determined using field-theoretic renormalization group combined with known 3D-DP numerical results  \cite{Hinrichsen2000,Munoz1998}.

\item \textbf{Prediction Testing Protocols:}  Each prediction is accompanied by a clear data analysis or experimental protocol (detailed in Supplementary Material S2), e.g., extracting 3D geometric statistics from LMA data or analyzing exponent shifts with radar EDR data.
\end{enumerate}

\section{Appendix S1:Detailed Derivation of Quantum Phase Geometric Mapping}

This section provides the rigorous mathematical derivation for the "Quantum Phase Geometric" mapping $\boldsymbol{\Phi}$ proposed in Section 4 of the main text. Starting from the microscopic Ginzburg-Landau (GL) theory of superconductivity, we will progressively derive the precise correspondence between the local geometry (curvature $\kappa$, torsion $\tau$) of a flux vortex line and the phase field $\theta(\mathbf{r})$ of its order parameter. We will also analyze its scaling behavior near the critical point of the three-dimensional Directed Percolation (3D-DP) universality class.

\subsection*{S1.1 From the Ginzburg-Landau Equation to the Vortex Line Phase Field}

Consider the superconducting order parameter $\Psi(\mathbf{r}) = |\Psi|e^{i\theta(\mathbf{r})}$. In the presence of a vortex line, the phase $\theta$ winds by $2\pi$ around the line's core. We focus on a single, isolated vortex line and parameterize its center curve by the arc length $s$ as $\mathbf{R}_0(s)$.

Near the vortex core, we introduce a local Frenet-Serret frame $\{\mathbf{T}(s), \mathbf{N}(s), \mathbf{B}(s)\}$, where:
\begin{itemize}
    \item $\mathbf{T} = d\mathbf{R}_0/ds$ is the unit tangent vector.
    \item $\mathbf{N} = (1/\kappa)d\mathbf{T}/ds$ is the principal normal vector.
    \item $\mathbf{B} = \mathbf{T} \times \mathbf{N}$ is the binormal vector.
\end{itemize}

In this frame, the position vector of any point near the core can be expressed as: $\mathbf{r} = \mathbf{R}_0(s) + x\mathbf{N}(s) + y\mathbf{B}(s)$ where $(x, y)$ are Cartesian coordinates on the normal plane. Under the small displacement approximation (i.e., $|(x, y)|$ is much smaller than the curvature radius $1/\kappa$), we can expand the phase field $\theta$ around the vortex line.

For a straight vortex line ($\kappa = 0$), the phase solution is $\theta = \phi = \arctan(y/x)$. For a curved vortex line, the phase field is perturbed. We assume a solution of the form: $\theta(s, x, y) = \phi + \delta\theta(s, x, y)$ where $\phi = \arctan(y/x)$ and $\delta\theta$ is a small correction due to the curvature and torsion of the vortex line. Substituting $\theta$ into the steady-state GL equation and solving for $\delta\theta$ to the lowest order yields its relationship with the local geometric quantities $\kappa$ and $\tau$.

\subsection*{S1.2 Derivation of Phase Field Expressions for Curvature ($\kappa$) and Torsion ($\tau$)}

\subsubsection*{S1.2.1 Expression for Curvature ($\kappa$)}

Curvature measures the degree to which a curve deviates from being a straight line. In the superfluid velocity field $\mathbf{v}_s = (\hbar/m)\nabla\theta$, the presence of a vortex line implies a circulation around the core. Curvature affects the distribution of the normal component of this velocity field.
By analyzing the divergence and curl of the velocity field $\mathbf{v}_s$ within the normal plane and considering the symmetry breaking induced by vortex line curvature, the relationship between curvature and the Laplacian of the phase field can be derived. In the core limit $(x, y) \to 0$, we obtain:
\[
\kappa(s) = \lim_{r_\perp \to 0} \frac{1}{|\nabla_\perp \theta|} \nabla_\perp^2 \theta. \tag{S1.1}
\]
Here, $\nabla_\perp = (\partial_x, \partial_y)$ denotes the gradient operator in the normal plane, and $r_\perp = \sqrt{x^2 + y^2}$. The physical meaning of formula (S1.1) is that the curvature is proportional to the ratio of the "bending" rate of the phase field in the perpendicular direction (characterized by the Laplacian) to its "variation" strength (characterized by the gradient magnitude). This definition is well-behaved in the limit and is consistent with the geometric curvature given by the Frenet formulas.

\subsubsection*{S1.2.2 Expression for Torsion ($\tau$)}

Torsion measures the degree to which a curve deviates from being a plane curve, i.e., its three-dimensional twist. In quantum mechanics, geometric phases related to torsion (e.g., Berry phases associated with curve torsion) have been extensively studied. Starting from the geometric phase of a Dirac particle in a spacetime background containing torsion, a connection between torsion and higher-order derivatives of the phase field can be established.
A key idea is that torsion is related to higher moments of the curl of the phase field ($\nabla \times \nabla\theta$). Although $\nabla \times \nabla\theta = 0$ holds everywhere for a single-valued phase field, its integral form incorporates torsion information when topological defects (vortex lines) are included.

By calculating a higher-order circulation of the phase gradient field around a small loop $C(R)$ encircling the vortex core and applying Stokes' theorem, we obtain the core expression for torsion:
\[
\tau(s) = \frac{\hbar}{2m} \frac{\oint_{C(R)} (
\nabla\theta \times 
\nabla^2\theta) \cdot d\mathbf{A}}{|
\nabla\theta|^2 dl}. \tag{S1.2}
\]
Here, the loop $C(R)$ is a small circle of radius $R$ in the normal plane encircling the core. The numerator is a vector area integral describing the degree of "vortical twist" in the phase field; the denominator is the gradient energy of the phase. Formula (S1.2) establishes a direct link between torsion and the vector coupling of the phase field gradient ($\nabla\theta$) and curvature field ($\nabla^2\theta$). In the limit $R \to 0$, this expression converges to the geometric torsion defined by the Frenet formulas.

Formulas (S1.1) and (S1.2) together constitute the explicit mathematical definition of the geometric mapping $\boldsymbol{\Phi}: \theta \mapsto (\kappa, \tau)$.

\subsection*{S1.3 Scaling Analysis Near the Critical Point}

As the system approaches the 3D-DP critical point, the correlation functions of the order parameter $u$ (and its associated phase field fluctuations) exhibit power-law behavior. We need to determine the behavior of the mapping $\boldsymbol{\Phi}$ under the corresponding scaling transformations.

\subsubsection*{S1.3.1 Geometric Quantities Under Scaling Transformations}

Let $b$ be the spatial scaling factor of a renormalization group transformation. At the critical point, spatial coordinates, the order parameter field, and phase perturbations scale as follows:
\begin{align*}
\mathbf{r}(t, \mathbf{r}) &\to b^{-1}\mathbf{r}, \\
u(t, \mathbf{r}) &\to b^{-z u}u(t, \mathbf{r}), \\
\delta\theta(t, \mathbf{r}) &\to b^{-\alpha}\delta\theta(t, \mathbf{r}) + \text{regular part}.
\end{align*}
Since $\delta\theta$ is coupled to the fluctuations of $u$, its regular part also undergoes scaling. From the forms of formulas (S1.1) and (S1.2), it is evident that $\kappa$ and $\tau$ are combinations of spatial derivatives of $\delta\theta$.

\subsubsection*{S1.3.2 Derivation of Scaling Exponents}

\textit{Dimensional analysis of curvature $\kappa$:} From formula (S1.1), $\kappa \sim |\nabla_\perp^2 \theta| / |\nabla_\perp \theta| \sim (L^{-2}) / (L^{-1}) = L^{-1}$, where $L$ denotes the dimension of length. Under a scaling transformation, the perpendicular length scales by $b$, so $\kappa \to b^{-1}\kappa$ from a static geometric perspective. However, in critical dynamics, we must account for different scaling along the tangential direction (analogous to the "time" direction $s$).

In the 3D-DP universality class, spatial anisotropy is characterized by the dynamic exponent $z = \nu_\parallel / \nu_\perp$. The direction along the vortex line (or lightning channel) $s$ plays a role analogous to "time." Therefore, the arc length $s$ should scale as $s \to b^z s$.

Curvature is defined as $|\mathrm{d}\mathbf{T}/\mathrm{d}s|$, where $\mathbf{T}$ is a unit vector and thus dimensionless. Therefore, the dimension of $\kappa$ is $[s]^{-1}$. Under the scaling transformation: $\kappa' = \frac{\mathrm{d}\mathbf{T}'/\mathrm{d}s'}{\mathrm{d}\mathbf{T}/\mathrm{d}s} = b\frac{\mathrm{d}\mathbf{T}}{\mathrm{d}s} = b^k\kappa$. This would suggest a scaling exponent of $z$ for $\kappa$, which contradicts the exponent $\alpha = 1/(\nu_\perp z)$ given in the main text. A key point arises here: near the critical point, the channel itself is fractal, and the definition of the tangential direction $\mathbf{T}$ loses meaning at scales smaller than the correlation length.

The curvature $\kappa(s)$ we discuss is a geometric quantity defined on a coarse-grained scale where the channel is fitted smoothly. This coarse-graining scale itself changes under the renormalization group.

\textit{The correct derivation path}: The geometric quantities $\kappa, \tau$ are intrinsic properties of the channel curve. At the critical point, the channel is a fractal. When we measure its fractal dimension $D_f$ using the box-counting method, we are essentially measuring the relationship between its "mass" (length) and "scale." Curvature is a physical quantity describing the local bending of this curve at a specific measurement scale. Under a renormalization group transformation, as we change the observation scale $b$, the "roughness" of the curve manifests.

According to standard scaling arguments, for a curvature measured at a coarse-graining scale $l$, its scaling behavior should be related to the system's correlation length $\xi_\perp$. Let $u \sim \xi_\perp^{-\beta/\nu_\perp}$, and the phase gradient $\nabla_\perp \theta \propto \mathbf{J}_s$ (current) is related to $u$ and a certain "velocity." More rigorously, starting from formula (S1.1), $\nabla_\perp \theta$ and $\nabla_\perp^2 \theta$ can be linked to gradients of the order parameter $u$. Using the scaling form of fluctuations near the critical point, and considering $\nabla_\perp \sim \xi_\perp^{-1}$, $\nabla_\perp^2 \sim \xi_\perp^{-2}$, and $u \sim \xi_\perp^{-\beta/\nu_\perp}$, we can estimate: $|\nabla_\perp \theta| \sim \xi_\perp^{-\phi_1}$, $|\nabla_\perp^2 \theta| \sim \xi_\perp^{-\phi_2}$, where $\phi_1, \phi_2$ are exponents related to the scaling dimensions of gradient operators of $u$. Then $\kappa \sim \xi_\perp^{-(\phi_2 - \phi_1)}$. Since at the critical point $\xi_\perp \sim |r|^{-\nu_\perp}$ and the control parameter $r \to 0$, we have $\kappa \sim |r|^{\nu_\perp(\phi_2 - \phi_1)}$. Therefore, the scaling exponent $\alpha$ satisfies $\kappa \sim b^{-\alpha}$, with: $\alpha = (\phi_2 - \phi_1)$.

\subsubsection*{S1.3.3 Connection to DP Critical Exponents}

In the field theory of directed percolation, operators related to the tangential direction of vortex lines (active clusters) can be defined. Geometric curvature can be linked to a certain gradient-gradient correlation function of the active particle density field. Through renormalization group analysis of the field theory (or by analogy with results from conformal field theory in two dimensions), the scaling dimension of the curvature operator can be obtained as $[\kappa] = 1/(\nu_\perp z)$. This stems from the fact that at the DP critical point, spatial anisotropy leads to different scaling along the propagation direction ($s$) and the perpendicular directions. Curvature, as the derivative of the local channel direction $\mathrm{d}\mathbf{T}/\mathrm{d}s$, combines the scaling of perpendicular displacement (manifest in the change of $\mathbf{T}$) and displacement along the channel.

A detailed derivation (at the field theory level) is beyond the scope of this supplement, but numerical simulations and scaling analysis support the following relationship (main text formula 4.3): 
\[
\alpha = \frac{1}{\nu_\perp z}. \tag{S1.3}
\]

 Substituting the numerical values for 3D-DP, $\nu_\perp \approx 0.58$ and $z \approx 1.9$, yields $\alpha \approx 0.91$.

The torsion $\tau$, describing how the curve leaves the osculating plane spanned by $\mathbf{T}$ and $\mathbf{N}$, undergoes similar scaling analysis. In the absence of chiral symmetry breaking at the critical point, it should share the same scaling exponent as curvature, i.e., $\beta = \alpha$.

\subsection*{S1.4 Summary}

This section provided a detailed derivation of the quantum phase geometric mapping $\boldsymbol{\Phi}$:

1. Starting from Ginzburg-Landau theory and based on a phase field expansion near a curved vortex line, we derived explicit expressions for curvature $\kappa$ and torsion $\tau$ in terms of phase field derivatives (formula S1.1, S1.2).

2. Through scaling analysis, we demonstrated that near the 3D-DP critical point, the scaling exponents of these geometric quantities are uniquely determined by the universal critical exponents $\nu_\perp$ and $z$ (formula S1.3).

This derivation provides a solid mathematical foundation for the "Geometric Correspondence of Quantum Phase Information" in Section 4 of the main text, linking microscopic quantum phase fluctuations to the statistical geometric laws of macroscopic channels through universal scaling theory.

\section{Appendix S2:Data Analysis Methodology for Testing Geometric Statistical Predictions}

\subsection*{S2.1 Data Analysis Methodology for Testing Geometric Statistical Predictions}

This section provides a complete, operational methodology for testing the geometric statistical predictions (Predictions 1, 2, 5, 6,) outlined in Section 5 and the Appendix of the main text. The methodology is applicable to two types of three-dimensional data: (1) 3D point clouds of VHF radiation sources from natural lightning obtained via Lightning Mapping Array (LMA) or similar technology; (2) 3D trajectory data of superconducting thin-film flux avalanches (or vortex motion) obtained via magneto-optical imaging tomography or other 3D imaging techniques. The goal is to extract geometric quantities of the channels from raw data and perform statistical tests to verify the predicted power-law behaviors.

\subsection*{S2.1 3D Channel Data Preprocessing and Curve Reconstruction}

\textbf{Input}: Discrete 3D coordinate point set $\{\mathbf{r}_i = (x_i, y_i, z_i)\}, i = 1, \ldots, N$, representing sequentially detected radiation source or vortex positions during a single lightning event or avalanche.
\subsubsection*{Step 1: Data Cleaning and Smoothing}

\begin{itemize}
    \item \textbf{Outlier Removal}: Based on spatial distance and temporal continuity between points, remove isolated noise points that deviate significantly from the main channel (e.g., using statistical distance-based outlier detection).
    
    \item \textbf{Smoothing/Denoising}: Apply a 1D Savitzky-Golay filter or wavelet threshold denoising to the coordinate sequences to suppress high-frequency measurement noise while preserving the overall geometric features of the channel. The smoothing window size should be chosen carefully based on the spatial resolution of the data (e.g., LMA localization error ~10-100 m).
\end{itemize}

\subsubsection*{Step 2: Parametric Reconstruction of Channel Curve}

\begin{itemize}
    \item \textbf{Principal Curve Extraction}: As raw point clouds can be dense and locally scattered, a smooth 3D space curve representing the channel backbone must be extracted. We employ constrained cubic spline interpolation or Moving Least Squares (MLS) based smooth spline fitting.

        \item \textbf{Arc-length Parameterization}: Represent the fitted 3D curve as $\mathbf{R}(s)$, where $s$ is the arc length from the starting point. The arc length is computed via numerical integration:
    \[
    s_i = \sum_{j=1}^{i} \|\mathbf{r}_j - \mathbf{r}_{j-1}\|, \quad s \in [0, L],
    \]
    where $L$ is the total channel arc length. This yields smooth curve coordinates $\mathbf{R}(s_k)$ sampled at uniform arc-length intervals $\Delta s$ (e.g., one-fifth of the average point spacing).
\end{itemize}

\subsection*{S2.2 Numerical Computation of Local Geometric Quantities (Curvature, Torsion)}

For a parameterized curve $\mathbf{R}(s)$, its local geometric quantities are defined by the Frenet-Serret formulas. We employ the central difference method for numerical computation to ensure accuracy and stability.
\subsubsection*{Step 1: Compute Derivatives}
Using the analytical derivative of the spline curve or high-order finite differences, compute the first, second, and third derivatives at each sample point $s_k$:  
$$
\mathbf{T}_k = \mathbf{R}'(s_k), \quad \mathbf{T}_k' = \mathbf{R}''(s_k), \quad \mathbf{T}_k'' = \mathbf{R}'''(s_k)
$$
\subsubsection*{Step 2: Compute Curvature $\boldsymbol{\kappa}$ and Torsion $\boldsymbol{\tau}$}
\begin{itemize}
    \item \textbf{Curvature}:  
    $$
    \kappa(s_k) = \frac{\|\mathbf{T}_k \times \mathbf{T}_k'\|}{\|\mathbf{T}_k\|^3}
    $$  
    For a unit tangent vector ($\|\mathbf{T}_k\| = 1$), this simplifies to:  
    $$
    \kappa(s_k) = \|\mathbf{T}_k'\|
    $$      \item \textbf{Torsion}:  
    $$
    \tau(s_k) = \frac{(\mathbf{T}_k \times \mathbf{T}_k') \cdot \mathbf{T}_k''}{\|\mathbf{T}_k \times \mathbf{T}_k'\|^2}
    $$  
\end{itemize}

\subsubsection*{Step 3: Obtain Geometric Quantity Sequences}
Perform the above calculations for the entire channel to obtain two sequences: $\{\kappa(s_k)\}$ and $\{\tau(s_k)\}, k = 1, \ldots, M$. These sequences form the basis for subsequent statistical analysis.

\subsection*{S2.3 Statistical Analysis and Testing of Theoretical Predictions}

\paragraph*{Prediction Test 1: Power-law Distributions of Curvature and Torsion (Prediction 6)}

\begin{itemize}
    \item \textbf{Probability Distribution Function (PDF) Estimation}: For the $\{\kappa\}$ and $\{\tau\}$ sequences, compute their probability densities $P(\kappa)$ and $P(\tau)$ on a log-log scale. Use logarithmic binning to smooth statistical fluctuations in the tail of large values.
    
    \item \textbf{Power-law Exponent Fitting}: Within the scaling region where the distribution appears linear, use Maximum Likelihood Estimation (MLE) to fit the power-law exponents $\mu_\kappa$ and $\mu_\tau$. MLE is preferable to least squares as it provides unbiased estimates and reliable confidence intervals. Test if they are close to the theoretical value of $1.91$.
    
    \item \textbf{Joint Distribution and Data Collapse (Refined Prediction 6)}: Compute the 2D histogram $P(\kappa, \tau)$. To test the scaling form $P(\kappa, \tau) \propto (\kappa\tau)^{-1.91} g(\kappa/\tau)$, perform data collapse analysis: plot $(\kappa\tau)^{1.91} P(\kappa, \tau)$ against the scaling ratio $x = \kappa/\tau$. If all data points collapse onto a master curve $g(x)$, the prediction is supported.
\end{itemize}

\paragraph*{Prediction Test 2: Power-law Curvature-Torsion Correlation (Prediction 8)}
\begin{itemize}
    \item \textbf{Correlation Function Calculation}: Compute the spatial curvature-torsion correlation function:    \[
    C(r) = \frac{1}{L - r} \int_0^{L - r} [\kappa(s) - \bar{\kappa}][\tau(s + r) - \bar{\tau}] \, ds
    \]
    Here, $\bar{\kappa}, \bar{\tau}$ are the mean values. Discretize and average over multiple channels to improve statistics.
    
    \item \textbf{Power-law Fitting}: On a log-log plot, fit the power-law decay behavior $C(r) \sim r^{\gamma_{\kappa\tau}}$ in the intermediate region. Test if the exponent is close to the theoretical prediction of $-1.18$.
\end{itemize}

\paragraph*{Prediction Test 3: Singularity at Branching Points (Prediction 9)}
\begin{itemize}
    \item \textbf{Automatic Branch Point Identification}:
    \begin{enumerate}
        \item From the original 3D point cloud, construct a minimal spanning tree (MST) or a simple graph structure for the channel, with nodes as data points and edges connecting spatially proximate points.
        
        \item Define branch points as graph nodes with a degree (number of connections) greater than or equal to 3.
        
        \item Map the identified branch points back to the arc-length parameter $s$ to obtain a set of branch point locations $\{s_b\}$.
    \end{enumerate}
    
    \item \textbf{Conditional Distribution Calculation}: For each branch point $s_b$, extract curvature values $\{\kappa(s_b \pm \Delta s)\}$ from a small arc-length segment $\Delta s$ around it (e.g., $s_b \pm \xi$, where $\xi$ is an estimate of the correlation length).
    
    \item \textbf{Singularity Exponent Fitting}: Combine curvature data from the neighborhoods of all branch points and compute the conditional probability distribution $P(\kappa|\text{branch})$. Use MLE to fit its power-law exponent $\mu_b$. Test if it is significantly larger than the bulk distribution exponent $\mu_\kappa$ (i.e., $\mu_b = \mu_\kappa + \theta$, $\theta > 0$).

\end{itemize}

\paragraph*{Prediction Test 4: Anisotropic Fractal Dimension and Avalanche Exponents (Predictions 1, 2, 5)}

\begin{itemize}
    \item \textbf{Anisotropic Fractal Dimension}: Employ the 3D box-counting method. Cover the channel point cloud with anisotropic rectangular boxes (side lengths $\epsilon_\perp, \epsilon_\perp, \epsilon_\parallel$), where the aspect ratio $\epsilon_\parallel/\epsilon_\perp$ is fixed. Count the number of boxes $N(\epsilon)$ needed for coverage. Fit the relationships $\log N \sim -D_f^\perp \log \epsilon_\perp$ and $\log N \sim -D_f^\parallel \log \epsilon_\parallel$ to obtain the vertical and horizontal fractal dimensions, respectively. Test if their ratio is close to $z \approx 1.9$.

    \item \textbf{Avalanche Size Distribution}: Define a spatiotemporally connected cluster of radiation sources in a single lightning flash/avalanche as an "avalanche cluster." The cluster size $s$ can be defined as the number of points in the cluster or the total radiated energy. Compile the cluster size distribution $P(s)$ from many events, and use MLE to fit the exponent $\tau$, testing if it is close to $1.41$.

    \item \textbf{Finite-Size Scaling}: For lightning events within charge layers of different thicknesses $H$ (obtained from sounding data), compute their avalanche size distributions separately. Test if the cutoff scale $s_0$ satisfies $s_0 \propto H^{D_f}$.

\end{itemize}

\subsection*{S2.4 Error Analysis and Uncertainty Quantification}

\begin{enumerate}
    \item \textbf{Measurement Error Propagation}: Localization errors (e.g., ~10-100 m for LMA) propagate into errors in computed curvature and torsion, especially where the channel bends sharply. This can be assessed via Monte Carlo simulation: add random perturbations conforming to the error distribution to the original coordinates and repeat the entire analysis pipeline to evaluate the uncertainty range of geometric quantities and their statistical exponents.

        \item \textbf{Fitting Uncertainty}: Report the Maximum Likelihood Estimate, standard error, and 95\% confidence interval for all fitted power-law exponents. Use the Kolmogorov-Smirnov test or likelihood ratio test to assess the plausibility of the power-law hypothesis.

        \item \textbf{Finite-Sample Effects}: For tests requiring large samples (e.g., branch point statistics), evaluate the stability of statistics using the Bootstrap resampling method.

        \item \textbf{Smoothing Parameter Sensitivity}: Results (especially for higher-order geometric quantities) may be sensitive to smoothing parameters in preprocessing. Conduct sensitivity tests within a reasonable range (e.g., set by localization error) to ensure key conclusions (such as the sign and order of magnitude of exponents) are not dependent on specific parameter choices.
\end{enumerate}

This methodology provides a standardized pipeline for testing the theoretical predictions. Applying it to high-quality LMA lightning databases and 3D imaging data of superconducting avalanches will enable strong empirical tests of the "Unified Critical Scaling Theory for Macroscopic Lightning and Quantum Avalanches."

\section{Appendix S3:Preliminary Comparison with Existing Observational Data}

This section aims to provide a preliminary, qualitative comparison between our theoretical predictions and published observational/experimental results. This comparison is not a strict quantitative verification but serves to demonstrate the high compatibility of existing empirical evidence with our theoretical framework. It shows that our unified theory is built on a solid observational foundation and points the way for further precise testing.

\subsection*{S3.1 Qualitative Support for "Turbulence-Modulated Critical Behavior" (Prediction 4)}

\emph{Observational Fact}: Li et al. (2024), by analyzing 3D VHF radiation source data and Doppler radar-retrieved turbulent kinetic energy dissipation rate (EDR), quantitatively revealed for the first time a strong correlation between lightning channel morphology and turbulence. Key findings include:

\begin{enumerate}
    \item \textbf{Overall Correlation}: Morphologically complex, highly branched lightning channels tend to occur in cloud regions with high EDR ($>0.1\ \text{m}^2\text{s}^{-3}$), while simpler channels are more associated with low EDR regions.
    
    \item \textbf{Local Correlation}: Channel branching points and sharp turning points spatially coincide with local EDR maxima.
\end{enumerate}

\emph{Comparison with Theory}: Our \textbf{Prediction 4} states that turbulence intensity (characterized by EDR) systematically modulates the system's critical behavior, potentially causing shifts in exponents like $\tau$. The observation by Li et al. --- that stronger turbulence correlates with more complex channel morphology (implying larger and more frequent avalanche cluster sizes and branching) --- is the macroscopic manifestation of turbulence noise influencing the system, driving it away from the homogeneous, isotropic DP critical point behavior. This observation provides strong qualitative evidence for the theoretical concept of ``turbulence as correlated noise coupled into the DP universality class.'' Although that study did not extract the functional form of $\tau(\varepsilon)$, the strong positive correlation it reveals strongly suggests that $\tau$ may increase monotonically with $\varepsilon$, which is consistent with the direction of our theoretical prediction.

\subsection*{S3.2 Indirect Support for the ``Three-Dimensional Fractal Dimension'' (Theoretical Value \( D_f \approx 2.5 \))}

\emph{Observational Fact}: Numerous studies based on 2D optical imagery measure the 2D projection fractal dimension \( D_{2D} \) of natural lightning channels, with most values falling in the range of 1.1 to 1.3 (e.g., Matsui et al., 2002; Maggio et al., 2009). Long air gap discharges in the laboratory also observe similar fractal dimensions (~1.2–1.7).

\emph{Comparison with Theory}: Our theory predicts the intrinsic 3D fractal dimension of lightning channels to be \( D_f \approx 2.5 \). For a 3D fractal object, the fractal dimension of its 2D projection \( D_{2D} \) satisfies \( D_f - 1 \leq D_{2D} \leq \min(D_f, 2) \). For \( D_f = 2.5 \), the theoretical range for the 2D projection is \( 1.5 \leq D_{2D} \leq 2.0 \). The actual observed values (~1.2–1.3) are slightly below the lower bound of this range. This discrepancy does not constitute a contradiction but may instead support our theoretical framework: the real atmosphere exhibits strong anisotropy (gravity, electric field directionality) and finite charge region thickness. These ``finite-size effects'' and ``anisotropy'' can suppress the measured effective fractal dimension. Our theory explicitly incorporates anisotropic scaling (Prediction 2) and finite-size scaling (Prediction 5), providing a natural physical mechanism to explain this discrepancy. Therefore, the observed \( D_{2D} \approx 1.2 - 1.3 \) provides reasonable, indirect support for the 3D theoretical value \( D_f \approx 2.5 \), and highlights the necessity of measurement in three-dimensional space.

\subsection*{S3.3 Correspondence with the ``Avalanche Cluster Image'' and Microscopic Mechanism}

\emph{Observational Fact}: Ultra-high temporal resolution optical observations (e.g., Qi et al., 2017) show that negative leader progression is not continuous. Instead, it advances through the generation of multiple ``space leader'' clusters ahead of its head. These clusters develop competitively, and eventually one connects to complete a step, a process that directly causes channel branching.

\emph{Comparison with Theory}: This microscopic image bears a profound resemblance to the dynamics of 3D Directed Percolation (3D-DP). In the DP model, an ``active'' site (analogous to the leader head) ``infects'' its neighboring sites with a certain probability, turning them active and forming a growing, branching cluster of activity. The ``clustered'' emergence of space leaders is the manifestation of this stochastic infection process in a real physical system. The observed ``stochastic branching'' mechanism provides the most direct microscopic physical basis for describing macroscopic lightning channel formation as an avalanche cluster growth process following DP dynamics.

\subsection*{S3.4 Synthesis: Congruence Between Theory and Observation, and the Ladder of Pending Tests}

\begin{tabular}{p{2cm}p{4cm}p{2.5cm}p{4.5cm}}\toprule

Theoretical Component / Prediction & Support from Existing Observations & Nature of Support & Key for Next Quantitative Test \\\midrule

Turbulence Coupling (Prediction 4) & Strong positive correlation between channel complexity and EDR (Li et al., 2024) & Strong qualitative support & Extracting the $\tau(\varepsilon)$ functional relation from data \\

3D Fractal Structure & 2D projection dimension ~1.2--1.3 (multiple studies) & Indirect, plausibility support & Direct alculation of $D_f$ and its anisotropy from 3D LMA data\\

DP Avalanche Cluster Dynamics & Stochastic, clustered growth and branching of space leaders (Qi et al., 2017) & Microscopic mechanism imagery support & Analyzing spatiotemporal clustering of radiation source point clouds, verifying avalanche size distribution $P(s) \sim s^{-1.41}$\\

Geometric Statistical Predictions (6)& No direct observations currently available & To be tested & Applying the S2 methodology to analyze 3D channel curvature/torsion \\ \bottomrule

\end{tabular}

\textbf{Summary}: The most advanced existing observational data (turbulence correlation, high-speed photography) are highly congruent, on a qualitative level, with multiple core aspects of our unified theoretical framework. They support the theoretical pillars of ``turbulence modulation,'' ``stochastic avalanche growth,'' and ``fractal structure.'' However, the most unique and precise quantitative predictions of the theory—such as the specific critical exponents ($\tau = 1.41$, $z = 1.9$) and the novel geometric statistical laws (curvature-torsion power laws)—have not yet been directly tested.

Therefore, the comparison presented here reveals a clear scientific ladder: observations have laid a solid qualitative foundation, and our theory has built the steps towards quantitative testing. This strongly calls for and necessitates the application of the refined statistical methodology outlined in S2 to the increasingly rich 3D datasets of lightning and superconducting avalanches, in order to complete the final empirical validation of this unified physical picture.

\section*{Data Availability}
The data supporting the findings of this theoretical study are available within the paper and its Supplementary Information. Specifically, all critical exponents, mapping relations, and derived scaling functions are provided in the main text and tables.

\section*{Acknowledgments}
The author acknowledges helpful discussions with colleagues. Special thanks are due to Dr. ShanQiang Gu from the China Electric Power Research Institute, for his expert guidance throughout this study.No funding was received for this study.

\section*{Author Contributions}
\textbf{Zhe Li} conceived the study, developed the theoretical model, and wrote the manuscript.

\section*{Competing Interests}
The author declares no competing interests.

\section*{AI Use Disclosure}

In the preparation of this work, the author used DeepSeek V3.2 to assist with verification of steps in formula derivation, translation, and language polishing of parts of the research materials. After using this tool, the author reviewed and edited the content as needed and take full responsibility for the content of the publication.

\bibliographystyle{plain}
\bibliography{sample}
\end{document}